\documentclass{iau_FM}
\usepackage{graphicx}

\title[The G4Jy Sample] 
{The MWA GLEAM 4-Jy (G4Jy) Sample} 

\author[Sarah V. White]   
{Sarah V. White (SVW)$^{1,2}$,
Thomas M.O. Franzen$^3$,
O. Ivy Wong$^4$,\newline
Anna D. Kapi{\'n}ska$^{5,4}$,
Chris Riseley$^6$,
Paul Hancock$^1$,
Joseph Callingham$^{3,7}$,
Richard Hunstead$^7$,
Natasha Hurley-Walker$^1$,
Chen Wu$^4$,
Nick Seymour$^1$, 
Jesse Swan$^8$,
Randall Wayth$^1$,
John S. Morgan$^1$,
Rajan Chhetri$^1$,
Carole Jackson$^{3,1}$,
Stuart Weston$^9$,
\and Tom Mauch$^2$}

\affiliation{
$^1$International Centre for Radio Astronomy Research (ICRAR),\\ Curtin University, Bentley, WA 6102, Australia \\ email: {\tt sarah.white@icrar.org} \\[\affilskip]
$^2$South African Radio Astronomy Observatory (SARAO), 2 Fir Street, \\ Black River Park, Observatory, 7925, South Africa\\
$^3$ASTRON, PO Box 2, 7990 AA Dwingeloo, The Netherlands\\
$^4$ICRAR, University of Western Australia M468, 35 Stirling Highway, WA 6009, Australia\\
$^5$National Radio Astronomy Observatory, 1003 Lopezville Rd, Socorro NM 87801, USA\\
$^6$CSIRO Astronomy and Space Science, PO Box 1130, Bentley, WA 6102, Australia\\
$^7$Sydney Institute for Astronomy (SIfA), University of Sydney, NSW 2006, Australia\\
$^8$School of Physical Sciences, University of Tasmania, Hobart, Tasmania, 7001 Australia\\
$^9$Institute for Radio Astronomy and Space Research (IRASR), Auckland University of Technology, Auckland 1010, New Zealand
}

\pubyear{2018}
\jname{Astronomy in Focus, Volume 1} 
\editors{Volker Beckmann \& Claudio Ricci, eds.}
\begin{document}
\maketitle

\begin{abstract}
Powerful radio-galaxies feature heavily in our understanding of galaxy evolution. However, when it comes to studying their properties as a function of redshift and/or environment, the most-detailed studies tend to be limited by small-number statistics. During Focus Meeting 3, on ``Radio Galaxies: Resolving the AGN phenomenon", SVW presented a new sample of nearly 2,000 of the brightest radio-sources in the southern sky (Dec. $<$ 30\,deg). These were observed at low radio-frequencies as part of the GaLactic and Extragalactic All-sky MWA (GLEAM) Survey, which is a continuum survey conducted using the Murchison Widefield Array (MWA). This instrument is the precursor telescope for the low-frequency component of the Square Kilometre Array, and allows us to select radio galaxies in an orientation-independent way (i.e. minimising the bias caused by Doppler boosting, inherent in high-frequency surveys). Being brighter than 4\,Jy at 151\,MHz, we refer to these objects as the GLEAM 4-Jy (G4Jy) Sample. The G4Jy catalogue is close to being finalised, with SVW describing how multi-wavelength data have been used to determine the morphology of the radio emission, and identify the host galaxy. In addition, the MWA's excellent spectral-coverage and sensitivity to extended/diffuse emission were highlighted. Both of these aspects are important for understanding the physical mechanisms that take place within active galaxies, and how they interact with their environment.

\keywords{galaxies: active, radio continuum: galaxies, galaxies: evolution, galaxies: jets.}
\end{abstract}

\firstsection 
\section{The brightest sources at low radio-frequencies}

Low-frequency radio observations allow us to probe older emission from an active galactic nucleus (AGN) than is traced at higher radio-frequencies, thereby providing information on the timescale over which an AGN may influence its host galaxy and surroundings. Moreover, this low-frequency emission is dominated by radio lobes. These are unaffected by Doppler boosting (also known as relativistic beaming), unlike the radio jets, hotspots, and core that dominate at high frequencies. As such, the angle at which we view the AGN has no bearing on the observed flux-density, and so selecting radio sources at low frequencies results in a sample that is free from orientation bias.  

Currently, the most prominent, low-frequency radio-source sample that is optically complete is the revised Third Cambridge Catalogue of Radio Sources (3CRR; \cite[Laing et al. 1983]{Laing83}). However, its flux-density limit (10\,Jy at 178\,MHz) restricts the detection of radio-bright galaxies to 173 sources. Requiring additional criteria, \cite[Wang \& Kaiser (2008)]{Wang2008} note that there is an insufficient number of objects for studying their cosmological evolution in detail, in terms of age or environmental density. A larger sample would also aid further investigation into the accretion modes of radio galaxies (\cite[Best \& Heckman 2012]{Best12}), and modelling of their dynamical behaviour (e.g. \cite[Turner \& Shabala 2015]{Turner15}). 

We create such a sample using new observations at low radio-frequencies, obtained via the Murchison Widefield Array (MWA). Thanks to its location in a protected, radio-quiet zone, we have excellent spectral-coverage, with 20 flux-density measurements spanning a frequency range of 72--231 MHz. In addition, the large number of short baselines ($<$\,200\,m) makes the MWA very sensitive to large-scale, diffuse radio emission. The data collected over the entire southern sky compose the GaLactic and Extragalactic All-sky MWA \cite[(GLEAM; Wayth et al. 2015)]{Wayth15} Survey, and the extragalactic component is publicly available in the form of a catalogue and images at multiple frequencies \cite[(Hurley-Walker et al. 2017)]{HurleyWalker17}\footnote{https://gleam-vo.icrar.org/gleam\_postage/q/form}. Using all radio sources with $S_{151\,\mathrm{MHz}} > 4$\,Jy in this catalogue, we construct the GLEAM 4-Jy Sample \cite[(Jackson et al. 2015)]{Jackson15}. This complete sample contains 1,860 sources and is over 10 times larger than 3CRR, due to its lower flux-density limit (4\,Jy at 151\,MHz) and larger survey area (24,831 square degrees). Like 3CRR, the majority of these sources are galaxies with an active black-hole at the centre, and many have radio jets associated with them. By using this larger sample to study radio-bright active galaxies, we can gain a better understanding of their fuelling mechanism, their connection with their environment, and how these radio sources evolve over cosmic time. 

\section{The G4Jy catalogue: determining the morphology and host galaxy}

\begin{figure}
\begin{center}
 \includegraphics[width=12.2cm]{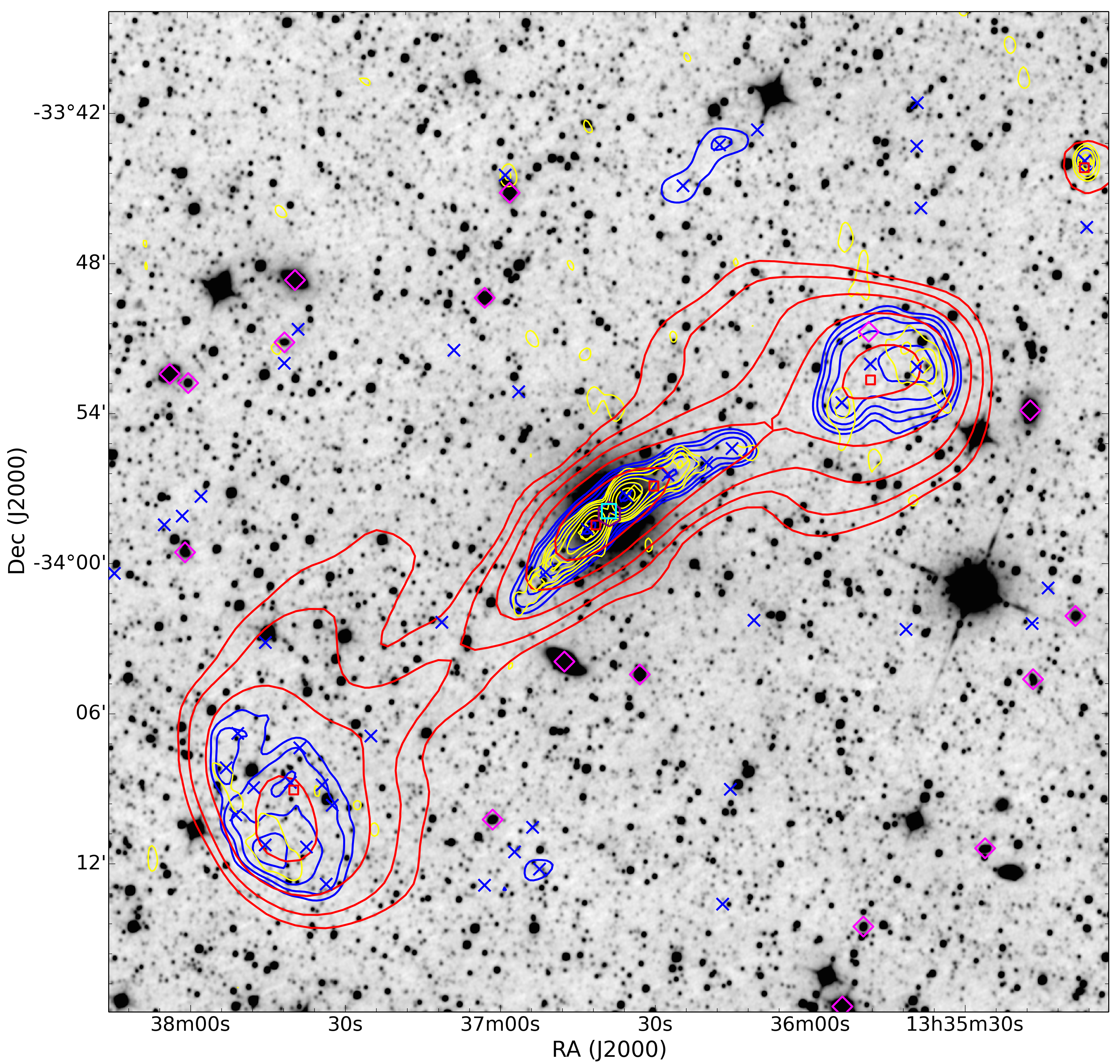} 
 \caption{An overlay, centred at R.A. = 13:36:39, Dec. = $-$33:57:57 (J2000), for an extended radio-galaxy (also known as IC~4296) in the G4Jy Sample. Radio contours from TGSS (150\,MHz; yellow), GLEAM (170--231\,MHz; red) and NVSS (1.4\,GHz; blue) are overlaid on a mid-infrared image from AllWISE ($3.4\,\mu$m; inverted grey-scale). For each set of contours, the lowest contour is at the 3\,$\sigma$ level (where $\sigma$ is the local rms), with the number of $\sigma$ doubling with each subsequent contour (i.e. 3, 6, 12\,$\sigma$, etc.). This source is an unusual example, in that its GLEAM-component positions (red squares) needed to be re-fitted using \textsc{Aegean} (\cite[Hancock et al. 2012]{Hancock12}, \cite[2018]{Hancock18}). Also plotted are catalogue positions from TGSS (yellow diamonds) and NVSS (blue crosses). The brightness-weighted centroid position, calculated using the NVSS components, is indicated by a purple hexagon.  The cyan square represents an AT20G (Australia Telescope 20-GHz Survey; \cite[Murphy et al. 2010]{Murphy10}) detection, marking the core of the radio galaxy. Magenta diamonds represent optical positions for sources in the 6-degree Field Galaxy Survey (\cite[Jones et al. 2004]{Jones2004}).}
   \label{fig1}
\end{center}
\end{figure}

The spatial resolution of MWA data ($\sim$2\,arcmin) means that we need to use other radio surveys for determining the morphology of the G4Jy sources. This involves visually inspecting overlays that use the TIFR GMRT Sky Survey (TGSS) alternative data release (ADR1; \cite[Intema et al. 2017]{Intema17}), providing 25-arcsec resolution, and either the NRAO VLA Sky Survey (NVSS; \cite[Condon et al. 1998]{Condon98}) or the Sydney University Molonglo Sky Survey (SUMSS; \cite[Mauch et al. 2003]{Mauch03}, \cite[Murphy et al. 2007]{Murphy07}), both of which provide 45-arcsec resolution. An example overlay is shown in Figure~\ref{fig1}, and demonstrates how only the GLEAM Survey reveals the extended distribution of relativistic plasma associated with the radio lobes. Spanning 40\,arcmin across, this radio galaxy is resolved into multiple GLEAM components by the MWA, as are a further 60 radio sources in our sample.

A crucial part of our careful visual inspection is to identify, where possible for each G4Jy source, the galaxy that hosts the radio emission. For this we use data from the mid-infrared AllWISE survey (\cite[Cutri et al. 2012]{Cutri12}), so that we are not biased against dust-obscured AGN. This is in contrast to many historical radio-source identifications, where radio contours were only overlaid onto {\it optical} images. Hence, we have completed thorough checks against the literature; reassessing `well-known' radio sources in light of relatively-new mid-infrared information, and searching for follow-up observations at high radio-frequencies (which may confirm the position of the radio core). The G4Jy catalogue will provide flags to highlight which sources would most-greatly benefit from further follow-up observations, due to there being multiple, candidate host-galaxies, or the AllWISE data having insufficient sensitivity. Meanwhile, our entire sample has been approved for follow-up by the Taipan Galaxy Survey (\cite[da Cunha et al. 2017]{daCunha17}), which is a new optical, spectroscopic survey over the southern hemisphere. 


\section{Constructing broadband radio spectra}

A great strength of the GLEAM Survey is the ability to tightly constrain the radio spectrum at low radio-frequencies. We extend this frequency coverage for a subset of the G4Jy Sample, by also incorporating measurements (at 5, 9, and 20\,GHz) from the Australia Telescope Compact Array. Doing so emphasises the degree of curvature in the radio spectrum, and that we cannot simply assume that radio emission follows a power-law description ($S_{\nu} \propto \nu^{\alpha}$, where $S$ is the flux density at a particular frequency, $\nu$, and $\alpha$ is the spectral index). Instead, there are poorly-studied physical processes that need to be considered. For `peaked-spectrum' sources (\cite[Callingham et al. 2017]{Callingham17}), which show `convex' spectra, synchrotron self-absorption or free-free absorption by ionised gas may be the cause of the spectral turnover towards low frequencies. As for sources that show an upturn or flattening in flux density towards high frequencies (i.e. have `concave' spectra), this may be due to renewed AGN activity or the orientation of the jet axis. At this point, we remind the reader that higher radio-frequencies probe more-recent/ongoing AGN activity, whilst observations at low frequencies reveal the cumulative effect of past epochs of activity. Therefore, spectral fitting allows constraints to be placed on the fraction of time that the central, supermassive black-hole is in active and quiescent phases (\cite[Turner~2018]{Turner18}) -- that is, the duty cycle of the AGN can be determined. 

\section{Summary and outlook}

The G4Jy Sample comprises the brightest radio-sources below Dec. $=30$\,deg, following repeated visual inspection of overlays and labour-intensive checks against the literature. Most importantly, the associated catalogue provides mid-infrared positions of $\sim$1,800 host galaxies, allowing the brightest detections in the GLEAM catalogue to be more easily cross-matched with information at other wavelengths. This will form a highly-valuable, legacy dataset in preparation for science enabled by the Square Kilometre Array. Furthermore, such a large, complete, unbiased sample is required for investigating what conditions promote, or are associated with, the production of powerful radio-jets. By exploiting the excellent spectral-coverage provided by the MWA, we can also tightly constrain the spectral behaviour of the G4Jy sources, and determine the prevalence of `restarted' AGN activity.

\end{document}